\documentclass[twoside,reqno]{heron}
\usepackage{epsfig,cite,colordvi}
\usepackage{graphicx}
\usepackage{url}
\usepackage{amssymb,amsmath,amscd,epsf}
\usepackage{times}
\usepackage{makeidx}
\begin{document}

\title{Pion-Nucleus Microscopic Optical Potential at Intermediate Energies and In-Medium Effect on the Elementary $\pi N$ Scattering Amplitude}

\runningheads{Pion-Nucleus Microscopic Optical Potential...}{Zemlyanaya, Lukyanov, Lukyanov, Zhabitskaya, Zhabitsky}

\begin{start}

\author{E.V. Zemlyanaya}{}, \coauthor{V.K. Lukyanov}{}, \coauthor{K.V. Lukyanov}{}, \\
\coauthor{E.I. Zhabitskaya}{}, \coauthor{M.V. Zhabitsky}{}

\index{Zemlyanaya, E.V.}
\index{Lukyanov, V.K.}
\index{Lukyanov, K.V.}
\index{Zhabitskaya, E.I.}
\index{Zhabitsky, M.V.}

\address{Joint Institute for Nuclear Research, 141980 Dubna, Russia}{}


\begin{Abstract}
Analysis is performed of calculations of the elastic scattering differential cross sections of pions on the $^{28}$Si, $^{40}$Ca, $^{58}$Ni and $^{208}$Pb nuclei at energies from 130  to 290 MeV basing on the microscopic optical potential (OP) constructed as an optical limit of a Glauber theory. Such an OP is defined by the corresponding target nucleus density distribution function and by the elementary  $\pi N$ amplitude of scattering.  The three (say, ``in-medium'') parameters of the $\pi N$ scattering amplitude: total cross section, the ratio of real to imaginary part of the forward $\pi N$ amplitude, and the slope parameter, were obtained by fitting them to the data on the respective pion-nucleus cross sections calculated by means of the corresponding relativistic wave equation with the above OP.
A difference is discussed between the best-fit ``in-medium'' parameters and the ``free'' parameters of the $\pi N$ scattering amplitudes known from the experimental data on scattering of pions on free nucleons.
\end{Abstract}
\end{start}

\section{Introduction}

There is a great number of papers on pion-nucleus scattering at different energies.
In theoretical study two approaches are usually employed.
{First}, the microscopic Kisslinger potential is based on $s-$, $p-$, and $d-$$\pi N$ scattering amplitudes having six and more parameters obtained from phase analysis of $\pi N$ data \cite{Kisslinger}.

{Second} approach is the Glauber high-energy approximation (HEA) that uses analytic form of the $\pi N$ amplitude inherent in high energy scattering \cite{Glauber}. Such approach was employed, for example, in \cite{ibraeva}.

Here we utilize our HEA-based microscopic optical potential (OP) \cite{ourpot} for calculation of $\pi$-nucleus elastic scattering.
This potential is constructed as an optical limit of a Glauber theory. Such an OP is defined by the known density distribution of a target nucleus and by the elementary  $\pi N$ amplitude of scattering.

The $\pi N$ amplitude itself depends on three parameters: total cross section $\sigma$, the ratio $\alpha$ of real to imaginary part of the forward scattering $\pi N$ amplitude, and the slope parameter $\beta$. For $\pi$-scattering on ``free'' nucleons they are known, in principle, from the phase analysis of the pion-nucleon  scattering data.  However, if one studies the pion-nucleus data then respective ``in-medium'' pion-nucleon amplitudes can be extracted. Thus the established best-fit ``in-medium'' $\pi N$ parameters can be compared with the corresponding  parameters of the ``free'' $\pi N$ scattering amplitudes.

The aim of our study is an explanation of the experimental pion-nucleus data in the region of (3~3)-resonance energies and estimation of the ``in-medium'' effect on the elementary pion-nucleon amplitude.

\section{Basic equations}
\label{model}

The differential cross sections are calculated by  solving the relativistic wave equation \cite{kaon} with the help of the standard DWUCK4 computer code \cite{dwuck}:
\begin{equation}\label{Shred}
  \left(\Delta + k^2 \right) \psi({\vec{r}})
  = 2 \bar\mu U (r) \psi({\vec{r}}),\quad U(r) = U^H(r) + U_C(r).
\end{equation}
Here $k$ is relativistic momentum of pion in c.m. system:
\begin{equation}
k={ \dfrac{M_A k^{\text{lab}}}{\sqrt{(M_A+m_\pi)^2+2M_A T^{\text{lab}}}}}\,,\quad
k^{\text{lab}}= \sqrt{ T ^{\text{lab}} \left( T^{\text{lab}} + 2m_{\pi}\right)}\,,
\end{equation}
$\bar\mu={E M_A}/[{E+M_A}]$ -- relativistic reduced mass, $E = \sqrt{k^2+m_\pi^2}$ -- total energy,
$m_\pi$ and $M_A$ -- the pion and nucleus masses,
$T^{\text{lab}}$ and $k^{\text{lab}}$ -- kinetic energy and momentum of pion in the laboratory system.

The HEA-based microscopic optical potential $U$ consists of nuclear and Cou\-lomb parts.
The nuclear part is as that derived in \cite{ourpot}:
 \begin{equation}\label{MOP}
     U^H  =  - \sigma \left(\alpha +i \right) \cdot
     \frac {{\hbar c} \beta_c}{ (2\pi)^2}
     \int_0^\infty
     {d q\,q^2 j_0(qr)\rho(q)f_{\pi}(q) }\,,
     \quad f_{\pi}(q)=e^{\tfrac{-\beta q^2}{2}}\,,
   \end{equation}
where
${\hbar c}=197.327$MeV$\cdot$fm,
$\beta_c={k}/{E},$
   $j_0$ is the spherical Bessel function,
$f_{\pi}(q)$
-- formfactor of $\pi N$-scattering amplitude,
$\rho(q)$ -- formfactor of the nuclear density distribution in the form of
symmetrized Fermi-function:
\begin{equation}\label{sf}
\rho_{SF}(r)=\rho_0\frac{\sinh\left( R/a \right)}
        {\cosh\left( R/a \right)+\cosh\left( r/a \right)},\, \rho_0={ \dfrac {A} {1.25\pi R^3} \left[1+(\frac{\pi a}{R})^2 \right]^{-1}}
        \end{equation}
Parameters of radius $R$ and diffuseness $a$ are known from electron-nucleus scattering data.

  Three parameters of the $\pi N$ scattering amplitude are obtained by fitting to the experimental $\pi A$ differential cross sections:
\begin{itemize}
\itemsep -1mm
\item
$\sigma$, total cross section $\pi N$,
\item
$\alpha$, ratio of real to imaginary part of the forward $\pi N$ amplitude,
\item
$\beta$, the slope parameter.
\end{itemize}
We minimize the function
\begin{equation}\label{Nev}
  \chi^2 =
  f \left( \sigma, \alpha, \beta \right)
  =\sum_{i}^{k}{ \frac{\left( y_i - \hat{y}_i(\sigma,\alpha, \beta)\right)^2}{({{s}^\text{as}}_{i})^2}},
\end{equation}
where  $y_i=\log {\displaystyle [\frac{d\sigma}{d\Omega}]_i}$ and $\hat{y_i}=\log {\displaystyle [\frac{d\sigma}{d\Omega}(\sigma,\alpha, \beta)]_i}$ are, respectively, experimental and
theoretical differential cross sections  of elastic scattering. Asymmetric experimental errors $s^\text{as}_i$ are  calculated at each $i$-th experimental point as follows
\begin{equation}\label{err}
  s^\text{as}_i=
        \begin{cases}
      y^{(+)}_i-y_i     \quad  \text{if} \ \hat{y}_i>y_i\ \, \\
      y_i-y^{(-)}_i      \quad \text{if} \ \hat{y}_i<y_i\ \,
    \end{cases}
\end{equation}
where $y^{(+)}_i$ and $y^{(-)}_i$ are, respectively, maximal and minimal estimations of the experimental value ${y}_i$.

The fitting technique is based on the asynchronous differential evolution algorithm \cite{ADE1,ADE2}.
\begin{center}
\begin{figure}[t]
\includegraphics[width=.32\linewidth,height= 6cm]{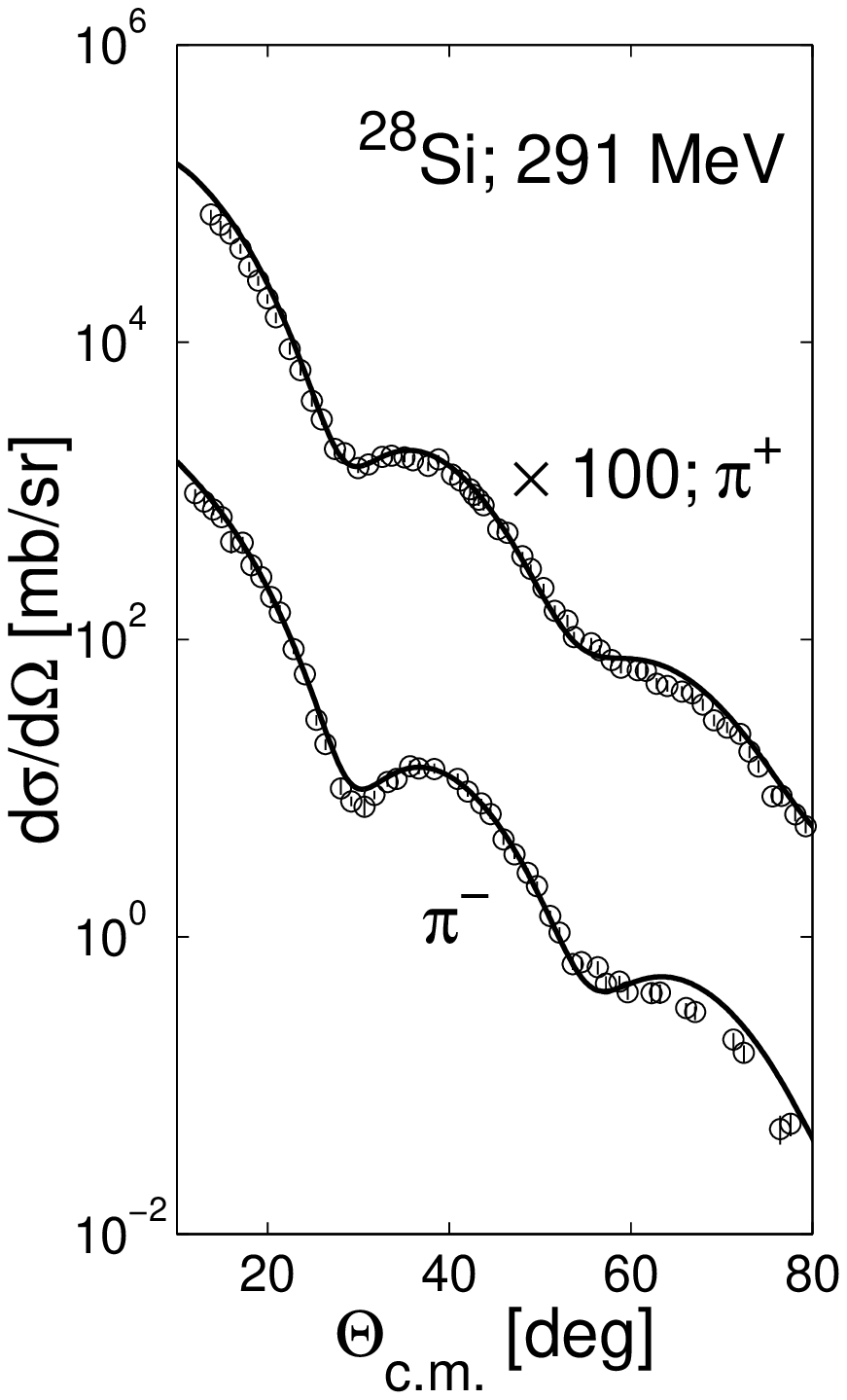}
\includegraphics[width=.32\linewidth, height= 6cm]{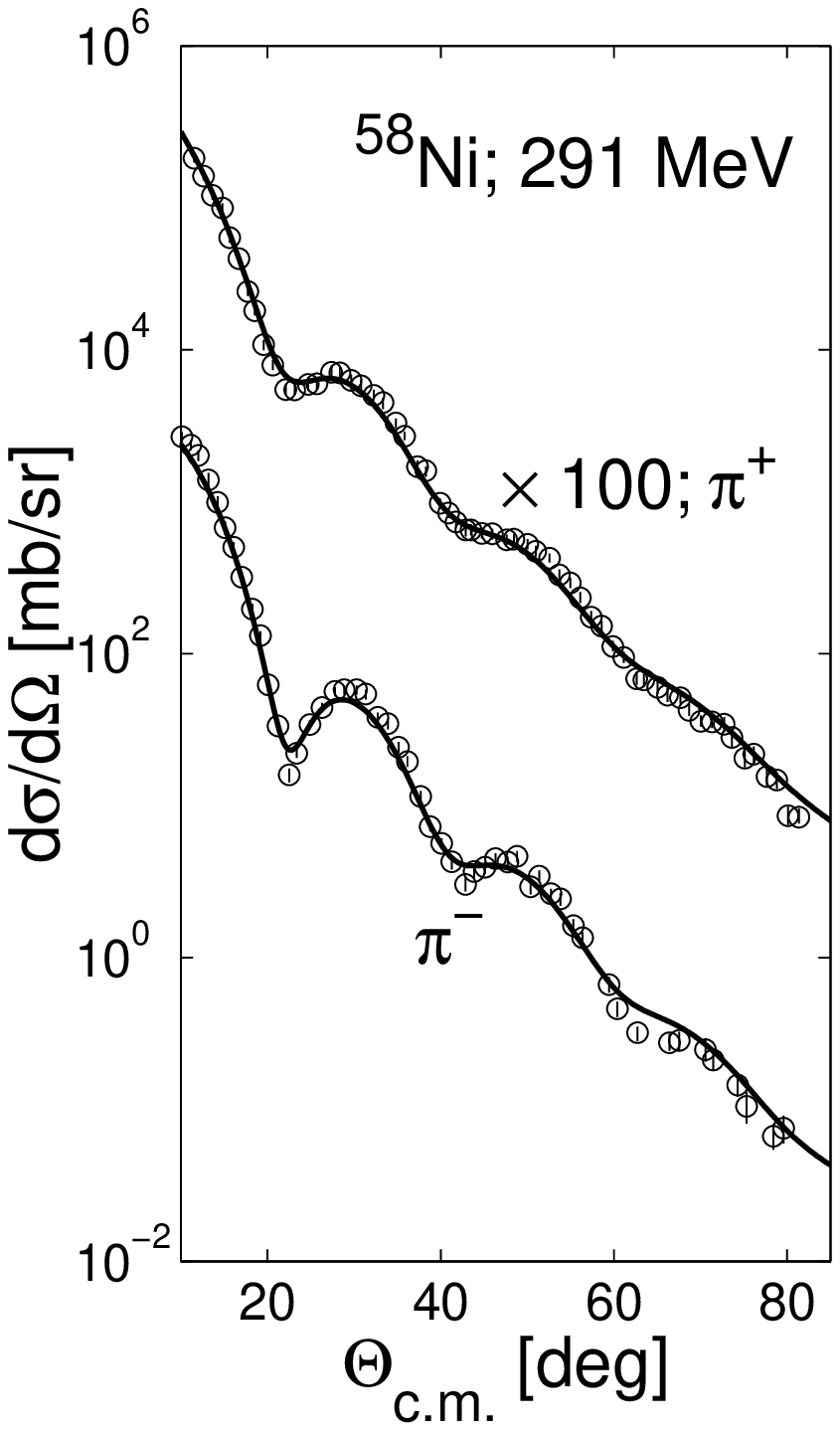}
\includegraphics[width=.32\linewidth, height= 6cm]{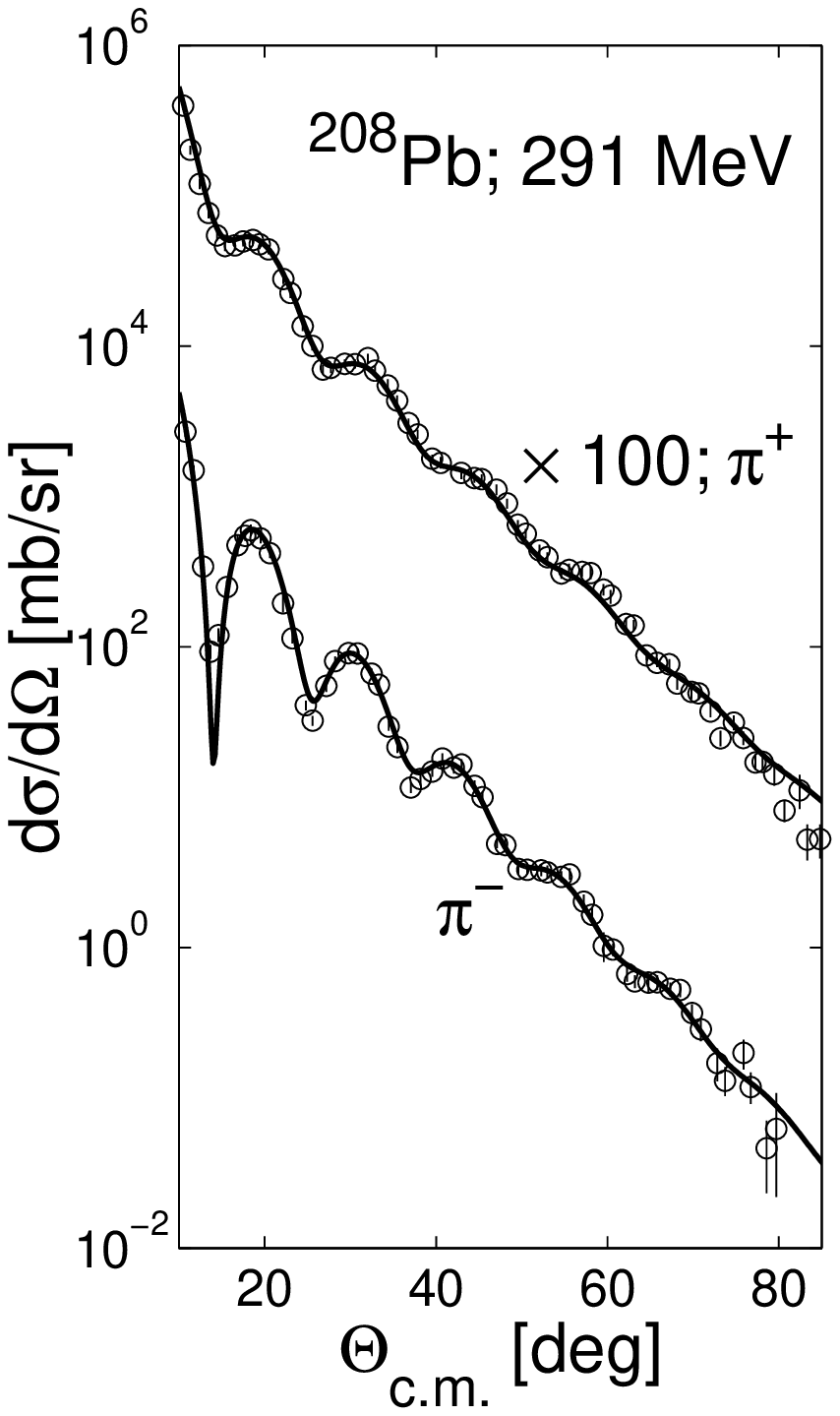}
\caption{Comparison of the calculated pion-nucleus elastic scattering differential cross sections at $T^{lab}=291$ MeV  with experimental data from \cite{Exper291}. The best-fit ``in-medium'' parameters $\sigma$, $\alpha$, and $\beta$ are given in the Table 1. \label{fig291}}
\end{figure}
\end{center}
\begin{center}
\begin{figure}[t]
\includegraphics[width=.32\linewidth, height= 6cm]{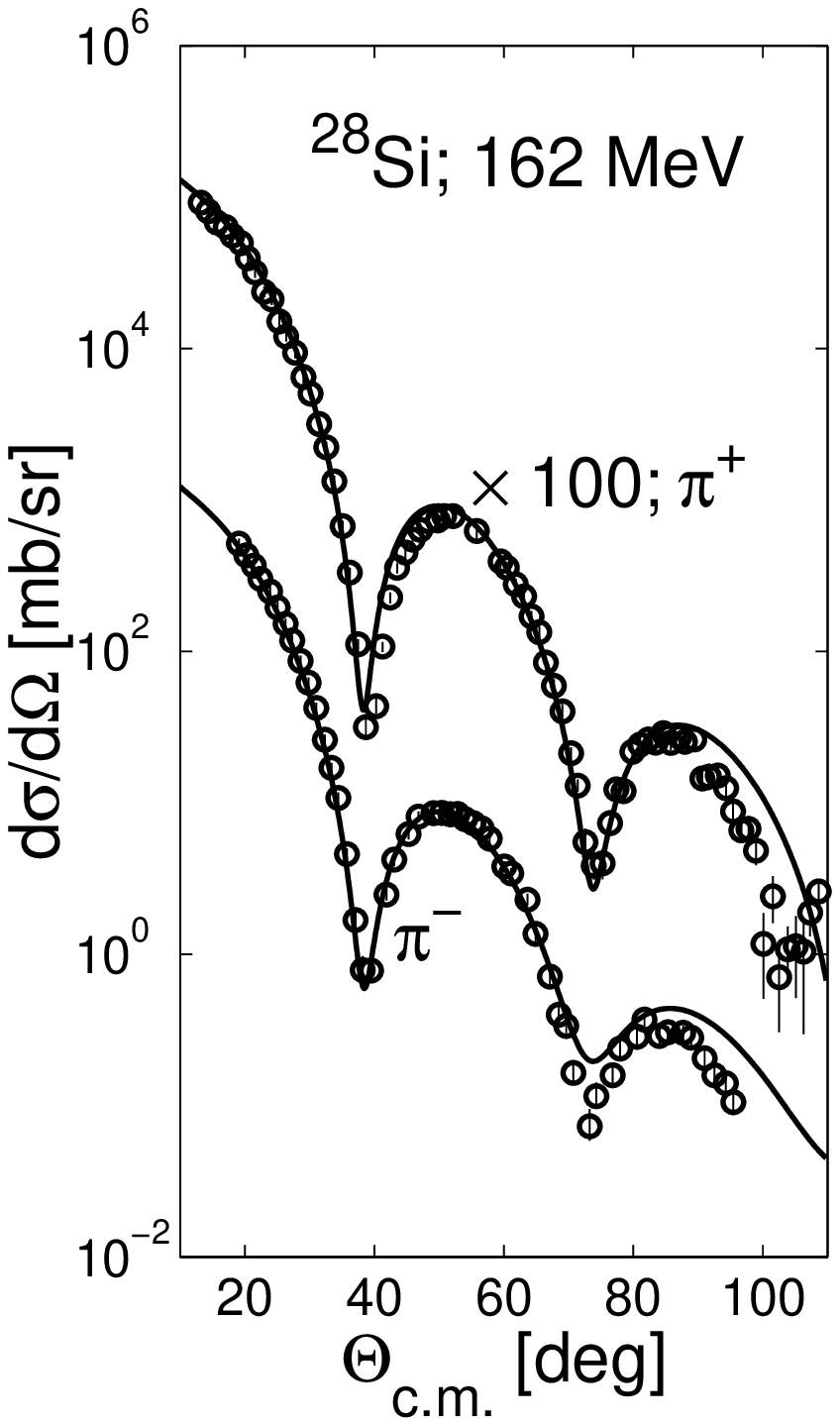}
\includegraphics[width=.32\linewidth, height= 6cm]{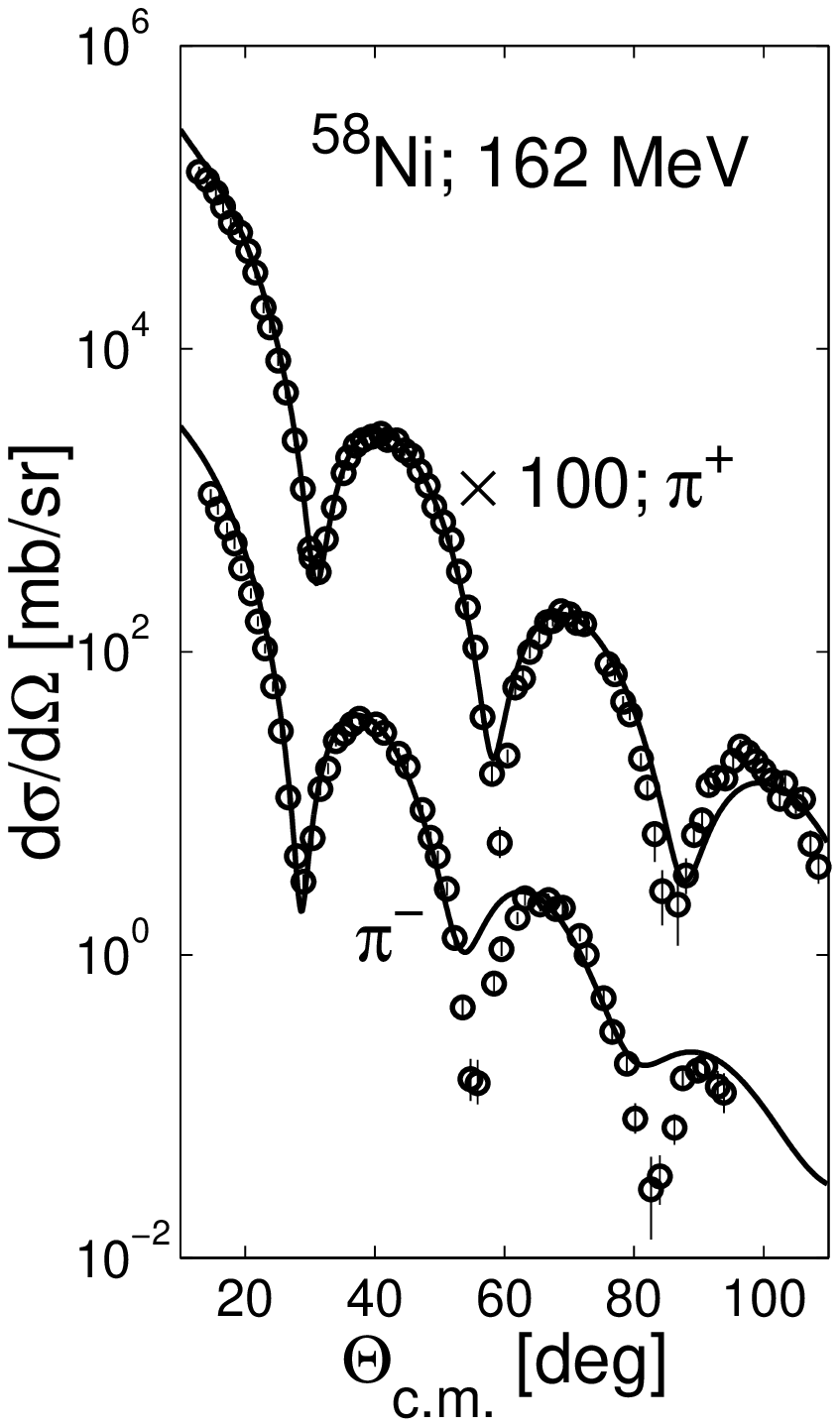}
\includegraphics[width=.32\linewidth, height= 6cm]{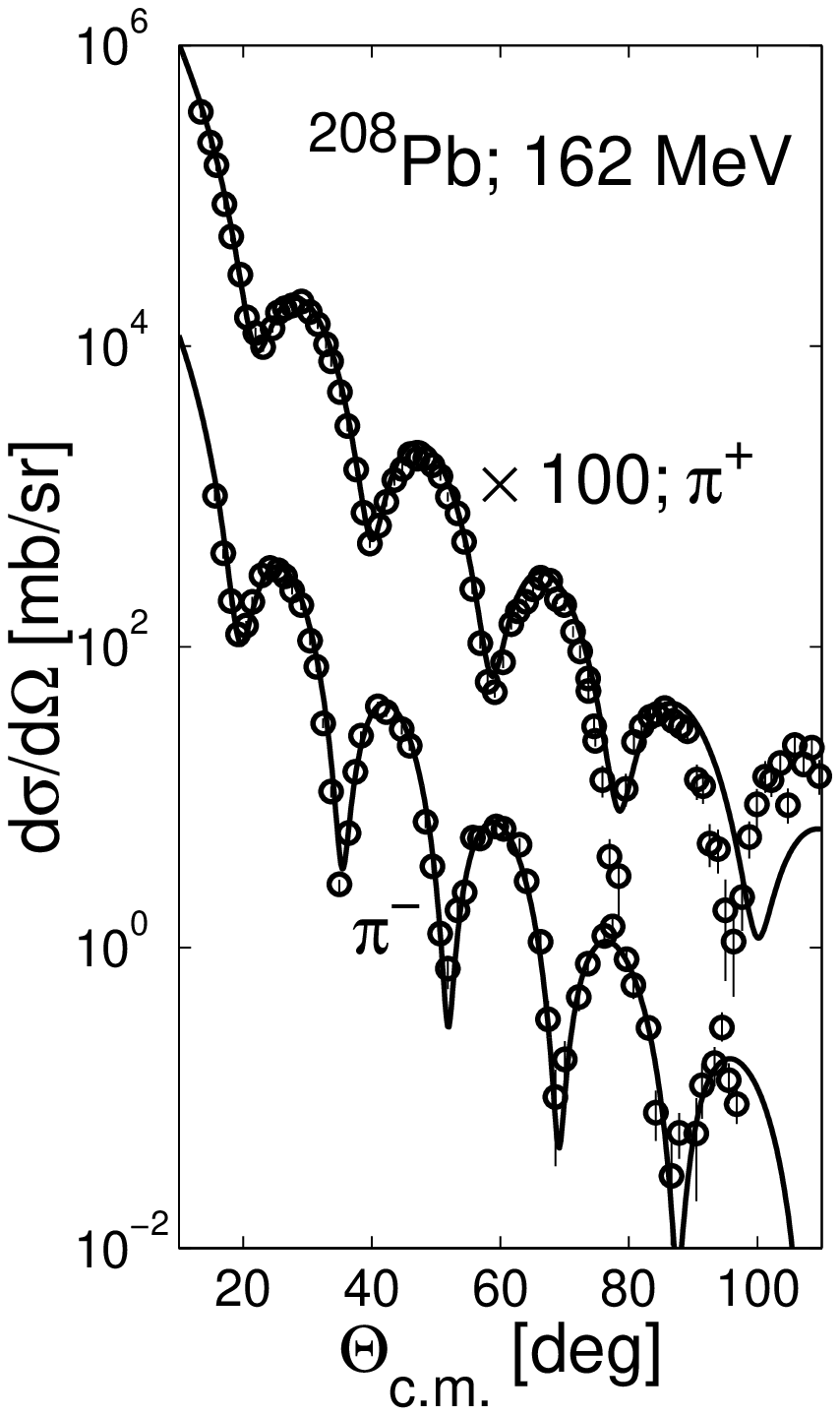}
\caption{The same as in Figure 1 but for $T^{lab}=162$ MeV. The experimental data are from \cite{Exper162}. \label{fig162}}
\end{figure}
\end{center}
\begin{center}
\begin{figure}[t]
\includegraphics[width=.45\linewidth, height= 8cm]{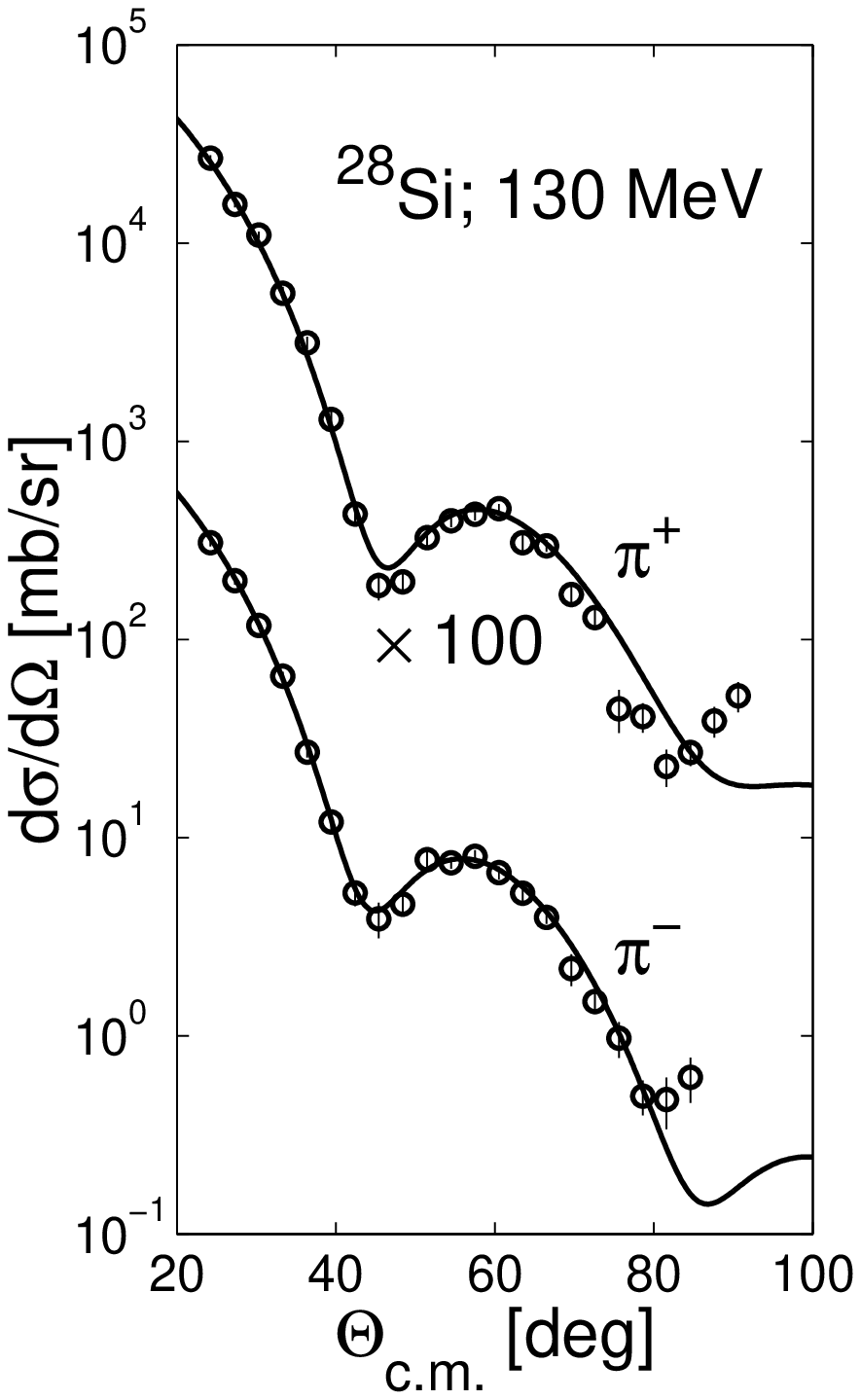}
\includegraphics[width=.45\linewidth, height= 8cm]{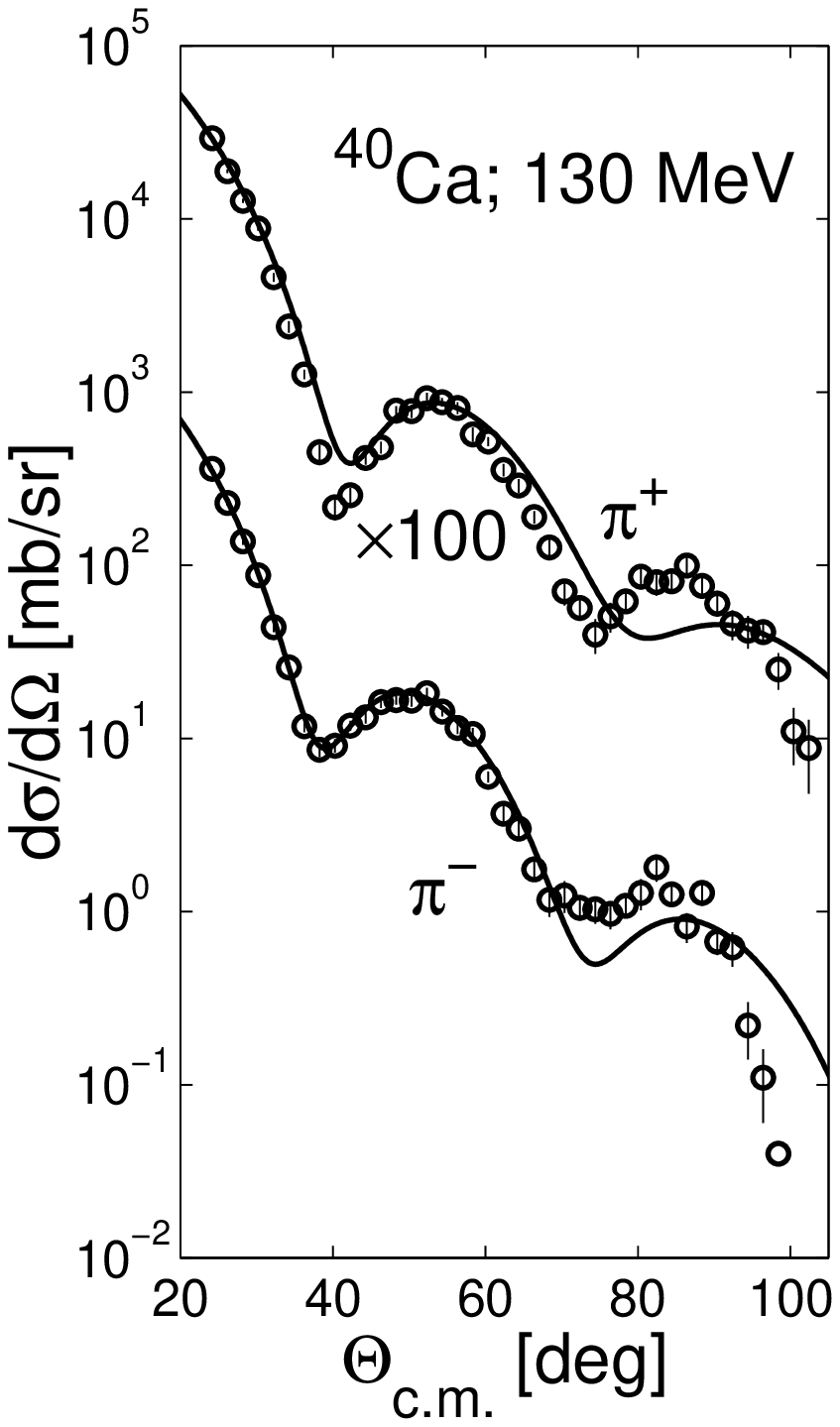}
\caption{The same as in Figure 1 but for $T^{lab}=130$ MeV. The experimental data are  from \cite{Exper_SI} and
 \cite{Exper_CA}. \label{fig130}}
\end{figure}
\end{center}
\begin{center}
\begin{figure}[t]
\includegraphics[width=.45\linewidth, height= 8cm]{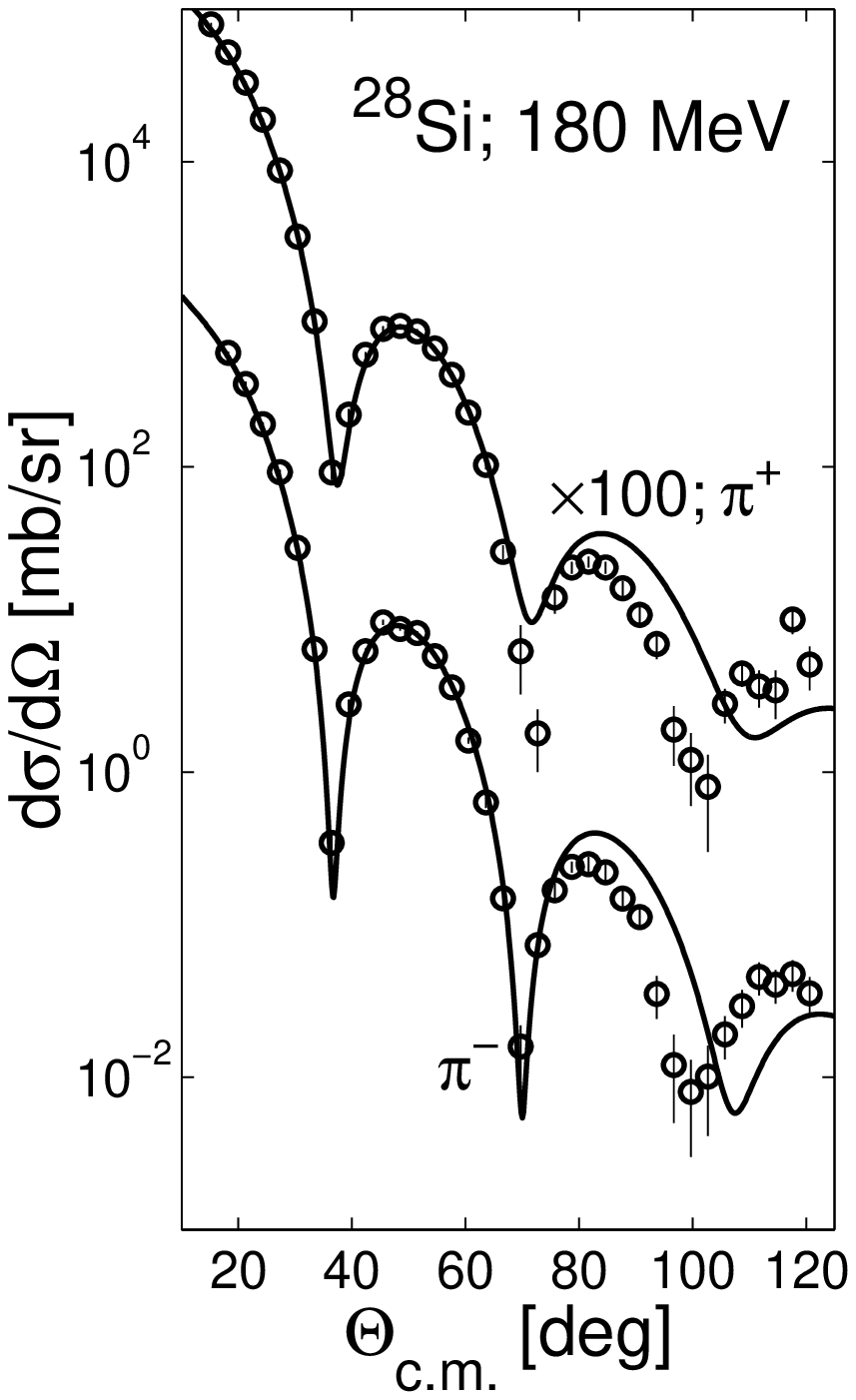}
\includegraphics[width=.45\linewidth, height= 8cm]{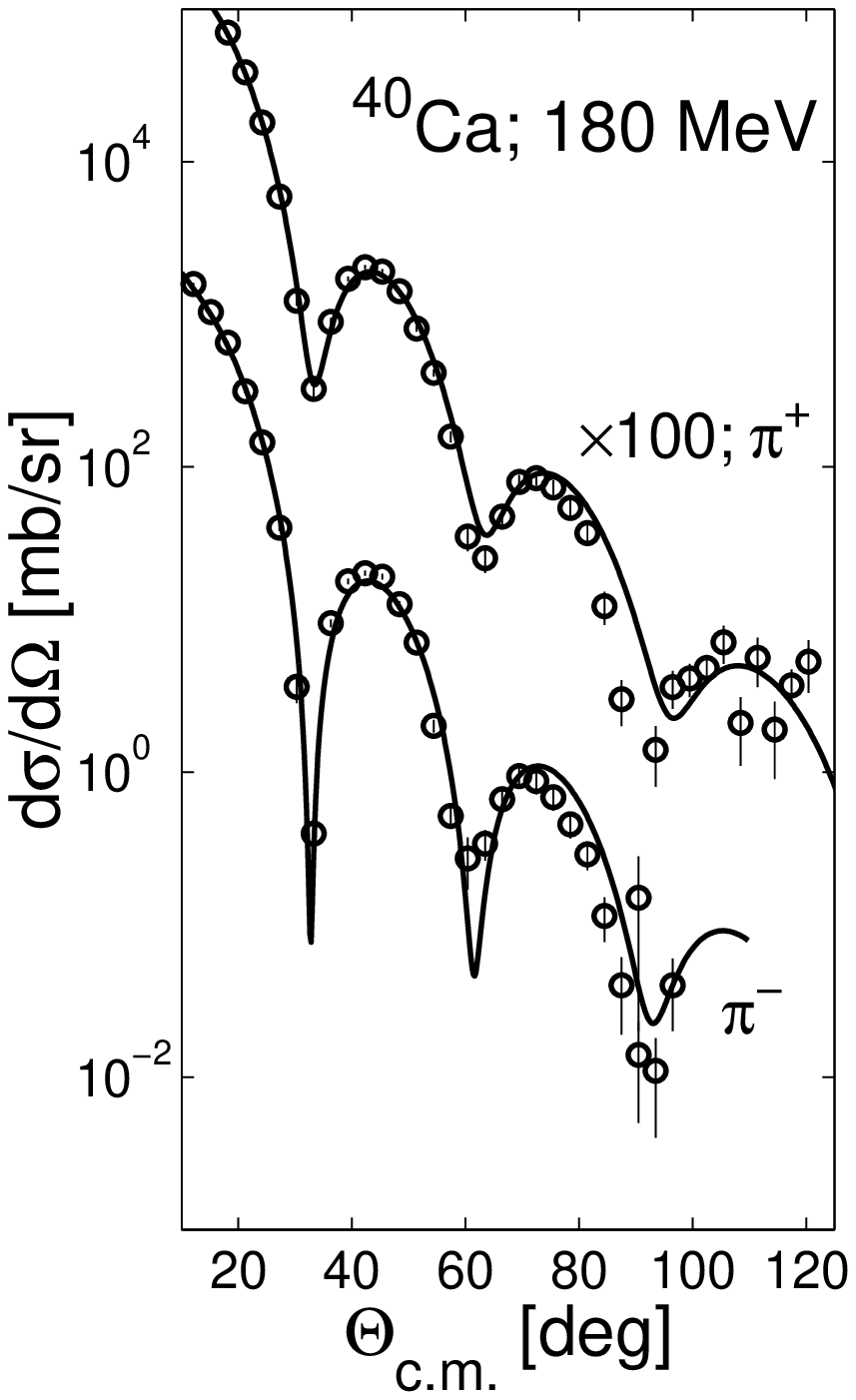}
\caption{The same as in Figure 1 but for $T^{lab}=180$ MeV. Experimantal data  are from \cite{Exper_SI} and
 \cite{Exper_CA}. \label{fig180}}
\end{figure}
\end{center}
\begin{center}
\begin{figure}[t]
\includegraphics[width=.45\linewidth, height= 8cm]{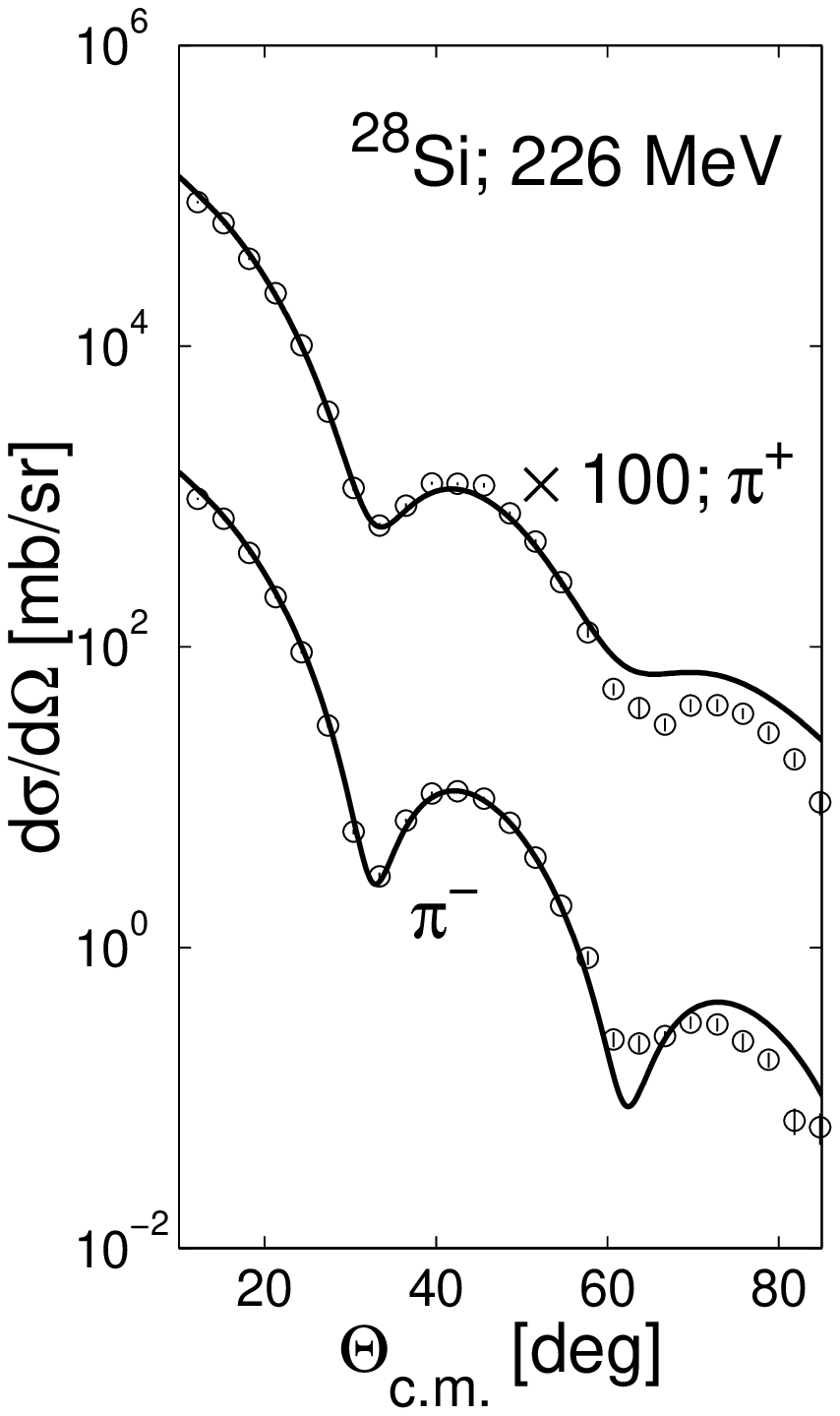}
\includegraphics[width=.46\linewidth, height= 8cm]{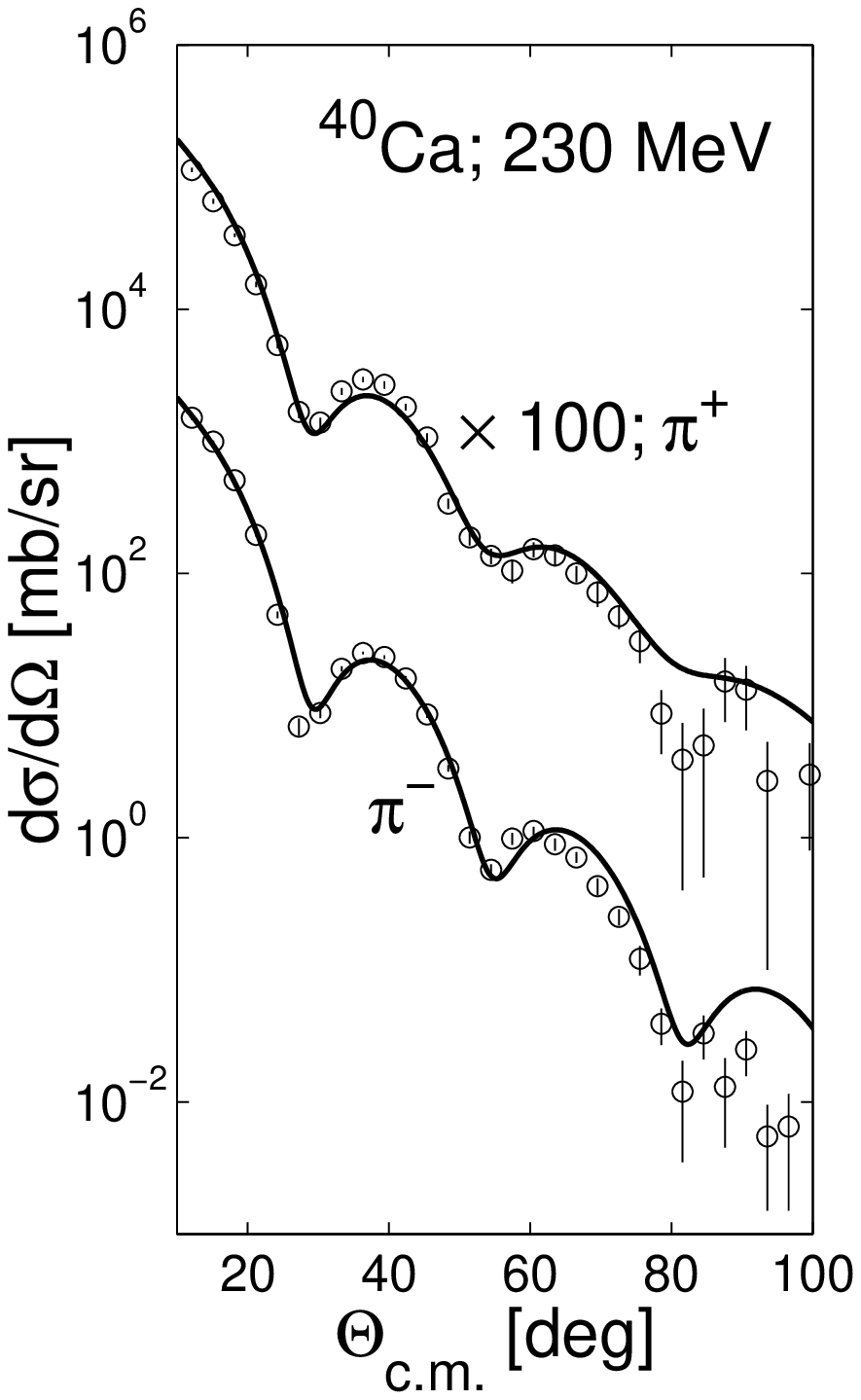}
\caption{The same as in Figure 1 but for $T^{lab}=226$ and 230 MeV.
Experimental data  are, respectively, from \cite{Exper_SI} and
 \cite{Exper_CA}. \label{fig226_230}}
\end{figure}
\end{center}
\begin{center}
\begin{figure}[t]
\begin{center}
\includegraphics[width=.6\linewidth]{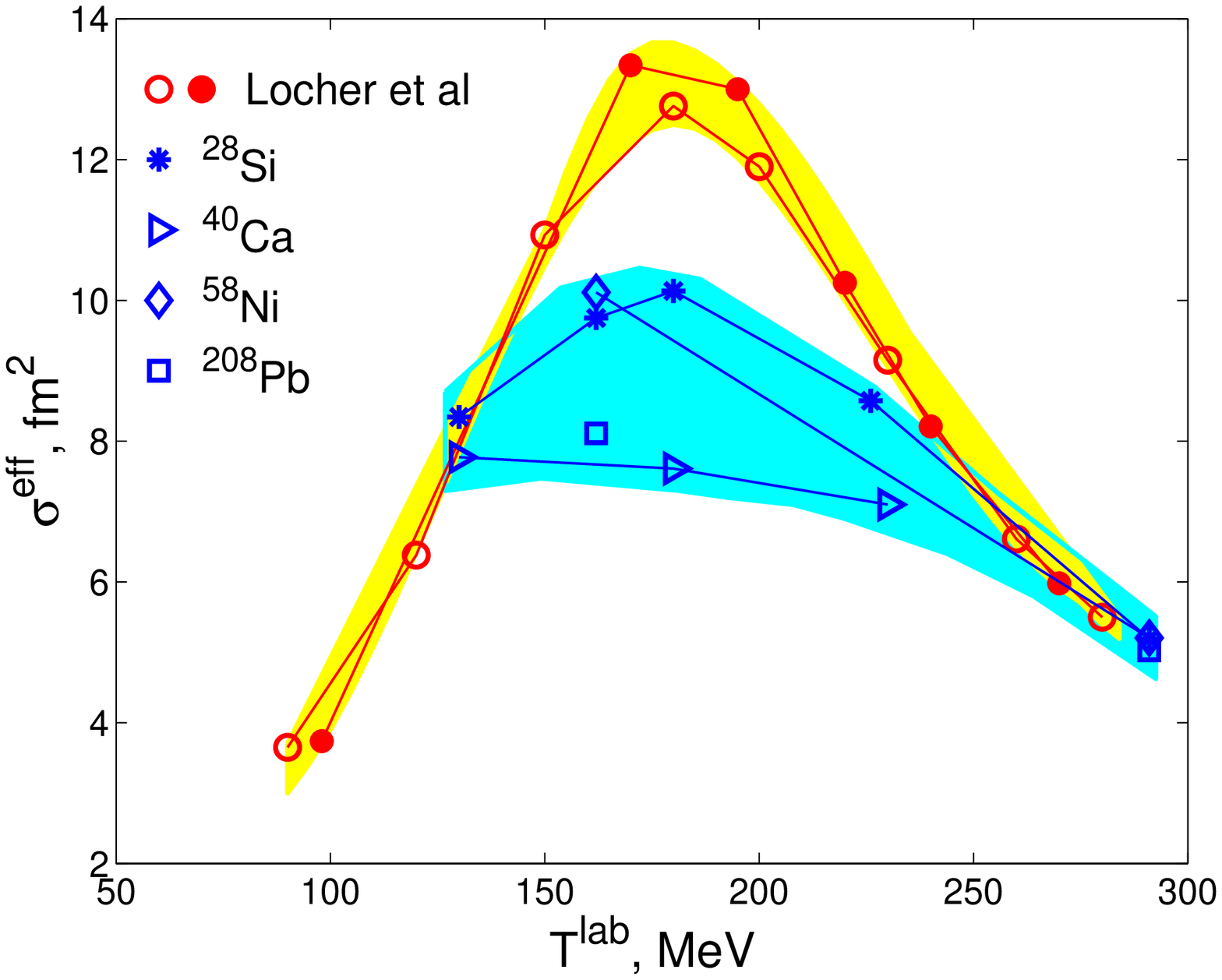}
\includegraphics[width=.6\linewidth]{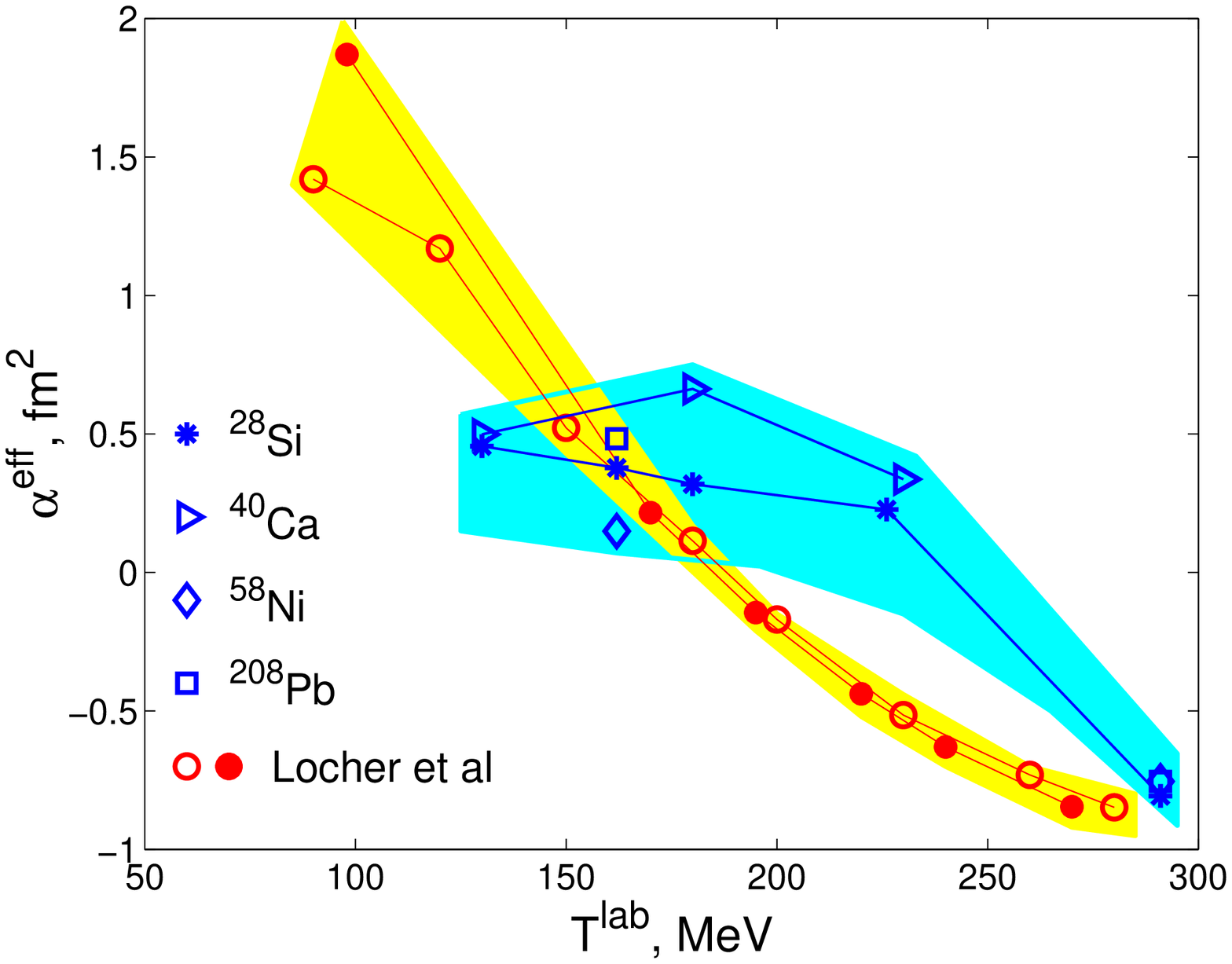}
\includegraphics[width=.6\linewidth]{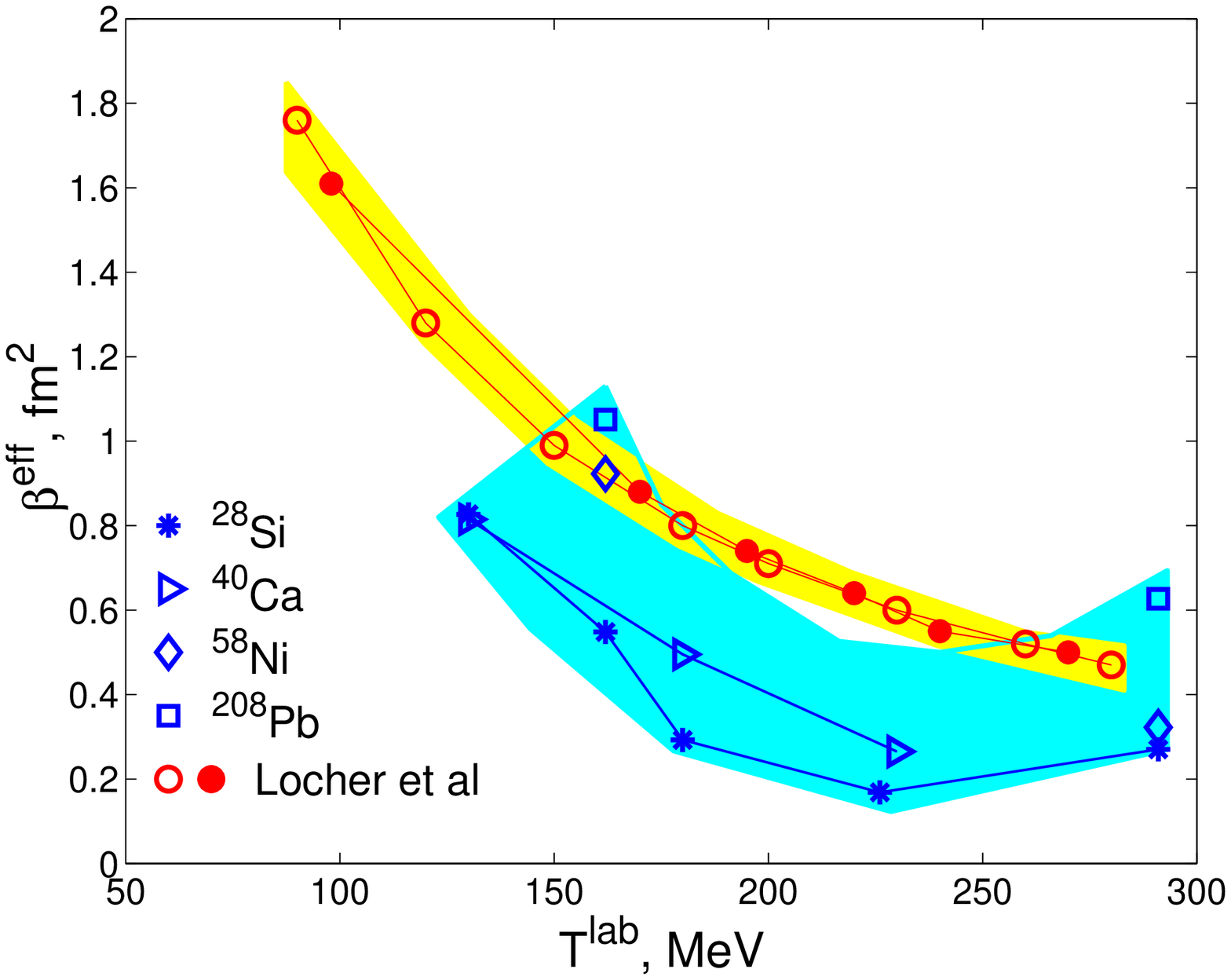}
\end{center}
\caption{(Color online) Light gray (yellow): ``free'' $\pi^{\pm}N$-scattering parameters from the paper of Locher {\it et al} \cite{Locher}. Dark gray (blue):  the best fit values $X^{eff} = (X_{\pi^+} + X_{\pi^-})/2$; $X=\sigma,\alpha,\beta$.\label{inmedium}}
\end{figure}
\end{center}
\begin{center}
\begin{figure}[t]
\includegraphics[width=.45\linewidth, height= 8cm]{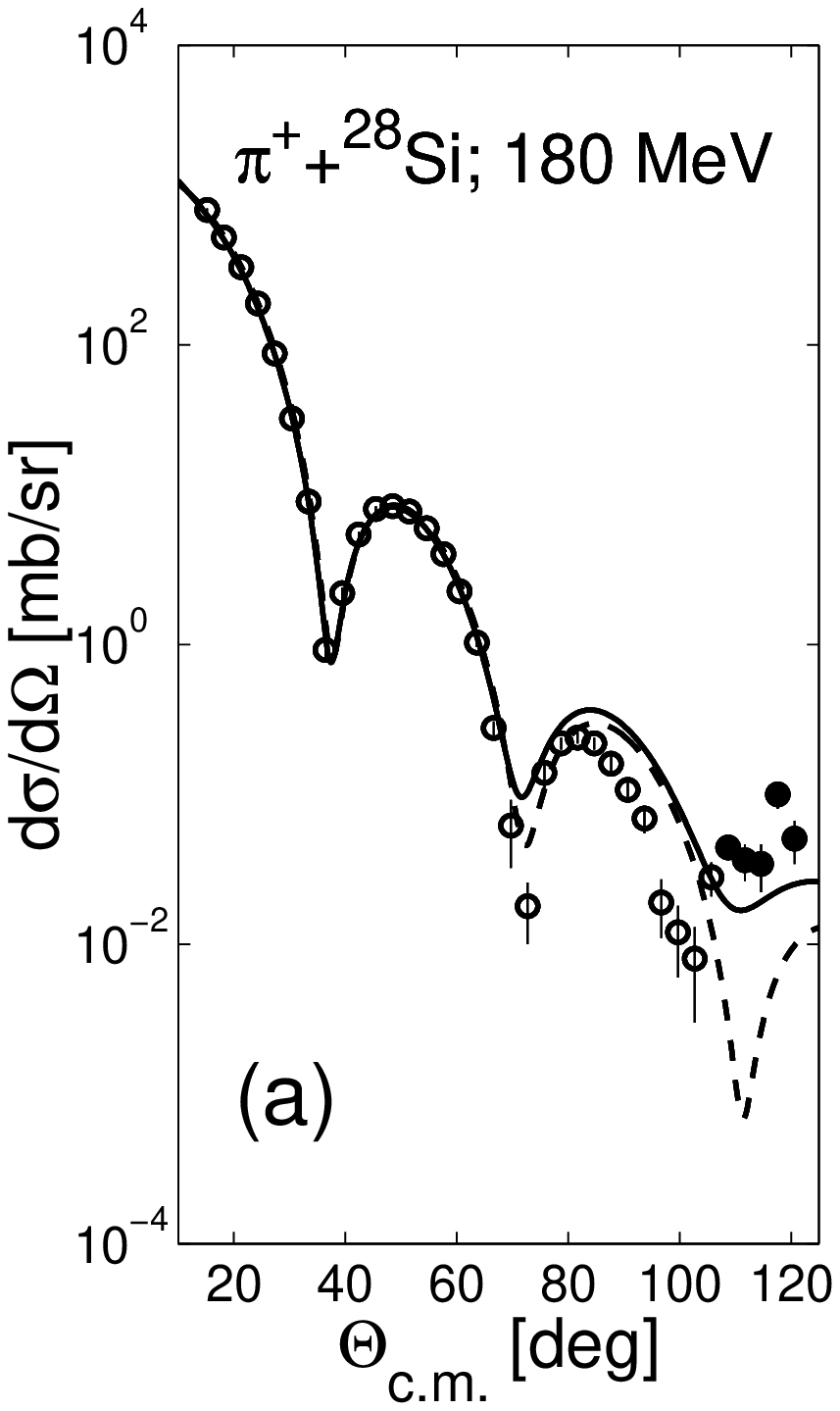}
\includegraphics[width=.46\linewidth, height= 8cm]{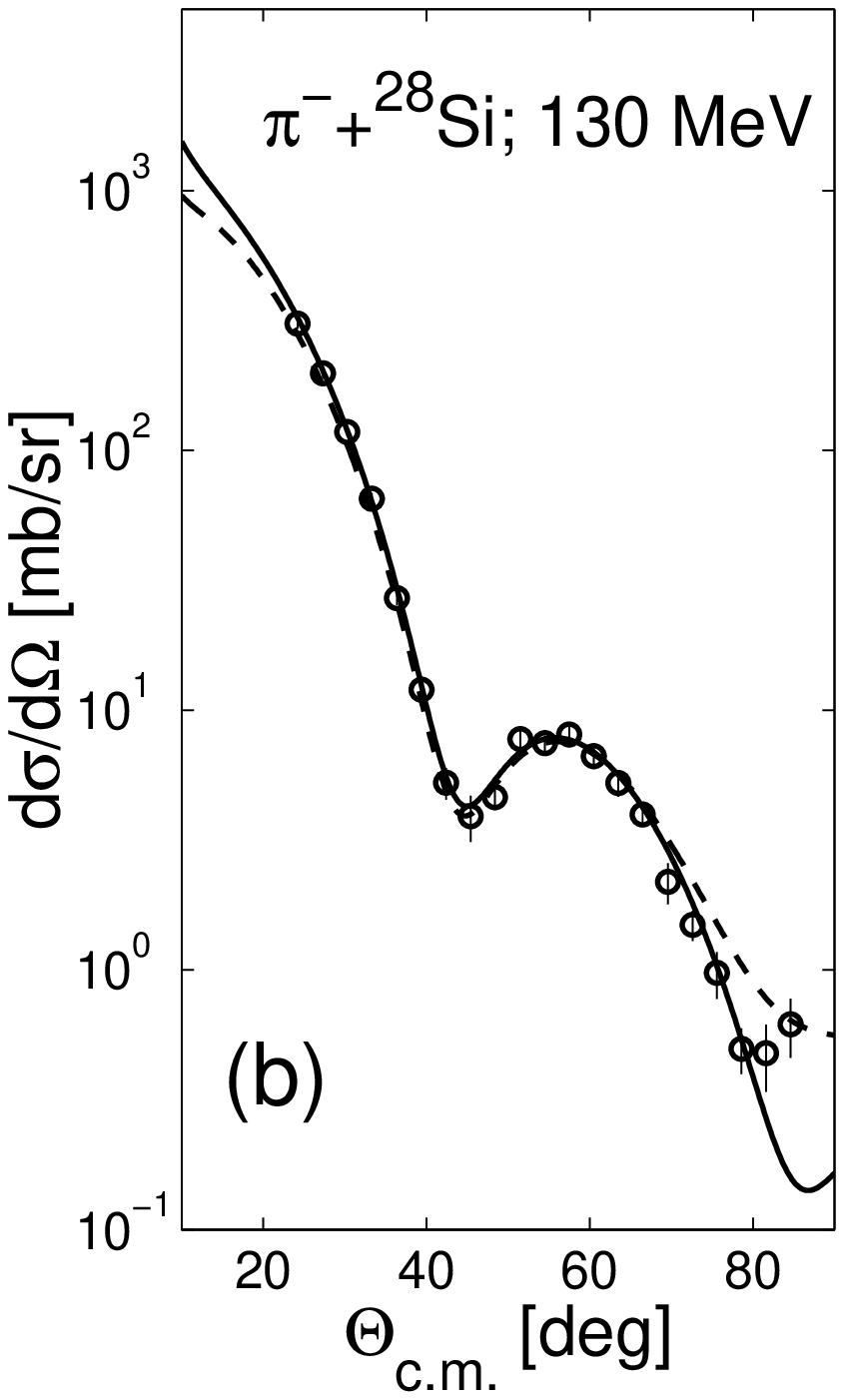}
\caption{(a) Differential cross sections of $\pi^++^{28}$Si scattering at 180 MeV. Solid curve: calculation with full set of experimental points from  \cite{Exper_SI}. Best-fit parameters are given in the Table 1. Dashed curve: calculation with reduced set of experimental points; the removed points are indicated by dark solid circles, and thus the obtained parameters are $\sigma=7.54$, $\alpha=0.771$, $\beta=.538$, $\chi^2/k = 5.1$.
      (b) Differential cross sections of $\pi^-+^{28}$Si elastic scattering at 130 MeV. Solid curve: calculation with the best-fit parameters from
      the Table 1. Dashed curve: calculation with parameters corresponding to the second minimum of the $\chi^2$ function where one gets  $\sigma=10.95$, $\alpha=-0.328$, $\beta=0.598$, $\chi^2/k=4.2$.  \label{fig80}}
\end{figure}
\end{center}
\begin{center}
\begin{table}[t]
\caption{The best-fit parameters $\sigma$, $\alpha$, $\beta$ and corresponding $\chi^2/k$ quantities where $k$ is the number of experimental points.}

\vspace*{3mm}

\begin{tabular}{l|c|c|l|l|l}
 \hline
  reaction&	$T^{lab}$ & $\chi^2/k$&  $\sigma$ &  $\alpha$ & $\beta$\\ \hline	
$\pi^-$+$^{28}$Si&130&2.1&7.08$\pm$0.16&0.87$\pm$0.05&0.81$\pm$0.05\\
$\pi^+$+$^{28}$Si&   &5.5&9.61$\pm$0.14&0.04$\pm$0.02&0.85$\pm$0.04\\

$\pi^-$+$^{40}$Ca&   &3.9 &6.97$\pm$0.11&0.89$\pm$0.01&0.87$\pm$0.03\\
$\pi^+$+$^{40}$Ca&   &13.3&8.58$\pm$0.08&0.11$\pm$0.01&0.76$\pm$0.02\\
\hline
$\pi^-$+$^{28}$Si&162&3.5&11.02$\pm$0.1&0.04$\pm$0.02&0.39$\pm$0.02\\
$\pi^+$+$^{28}$Si&   &6.7&8.48$\pm$0.06&0.71$\pm$0.01&0.71$\pm$0.01\\
$\pi^-$+$^{58}$Ni&   &10.7&10.95$\pm$0.1&-0.146$\pm$0.01&1.08$\pm$0.02\\
$\pi^+$+$^{58}$Ni&   &7.5 &9.28$\pm$0.04&0.444$\pm$0.01&0.77$\pm$0.01\\
$\pi^-$+$^{208}$Pb&  &3.7 &9.62$\pm$0.09&0.36$\pm$0.01&1.02$\pm$0.01\\
$\pi^+$+$^{208}$Pb&  &10.3&6.60$\pm$0.03&0.61$\pm$0.01&0.01$\pm$0.01\\
\hline
$\pi^-$+$^{28}$Si&180&10.5&10.03$\pm$0.06&0.33$\pm$0.01&0.266$\pm$0.01\\
$\pi^+$+$^{28}$Si&   &12.1&10.24$\pm$0.07&0.31$\pm$0.01&0.323$\pm$0.01\\
$\pi^-$+$^{40}$Ca&   &3.3&9.44$\pm$0.11&0.25$\pm$0.02&0.29$\pm$0.01\\
$\pi^+$+$^{40}$Ca&   &4.2&5.78$\pm$0.07&1.08$\pm$0.02&0.70$\pm$0.02\\
\hline
$\pi^-$+$^{28}$Si&226&13.8&7.36$\pm$0.06&0.596$\pm$0.01&0.175$\pm$0.01\\
$\pi^+$+$^{28}$Si&   &23.8&9.79$\pm$0.014&-0.142$\pm$0.02&0.162$\pm$0.013\\
\hline
$\pi^-$+$^{40}$Ca&230&7.56 &5.25$\pm$0.06&0.796$\pm$0.01&0.253$\pm$0.01\\
$\pi^+$+$^{40}$Ca&   &7.70 &8.95$\pm$0.02&-0.122$\pm$0.02&0.277$\pm$0.01\\
\hline
$\pi^-$+$^{28}$Si&291&6.2&5.03$\pm$0.08&-0.82$\pm$0.02&0.173$\pm$0.012\\
$\pi^+$+$^{28}$Si&   &4.9&5.35$\pm$0.13&-0.79$\pm$0.02&0.38$\pm$0.013\\
$\pi^-$+$^{58}$Ni&   &3.8&4.78$\pm$0.08&-0.85$\pm$0.02&0.28$\pm$0.02\\
$\pi^+$+$^{58}$Ni&   &2.6&5.63$\pm$0.15&-0.66$\pm$0.02&0.36$\pm$0.01\\
$\pi^-$+$^{208}$Pb&  &4.1&4.50$\pm$0.07&-1.06$\pm$0.02&0.666$\pm$0.02\\
$\pi^+$+$^{208}$Pb&  &3.0&5.56$\pm$0.15&-0.45$\pm$0.02&0.588$\pm$0.02\\
\hline
\end{tabular}
\end{table}
\end{center}

\section{Results and discussion}

Figures 1--5 show the differential cross  sections of elastic pion-nucleus scattering at energies between 291 and 130 MeV calculated using the OP  presented in Section \ref{model}. Parameters (radius $R$ and diffuseness $a$) of the target nuclear density distribution are following:
$R=3.134$ fm and $a=0.477$ fm for ${^{28}}$Si \cite{LZS2004};
$R=4.2$ fm and $a=0.475$ fm for ${^{58}}$Ni \cite{ElAzab1979};
$R= 3.593$ fm and $a=0.493$ fm for $^{40}$Ca  \cite{LZS2004};
$R=6.654$ fm and $a=0.475$ fm for ${^{208}}$Pb \cite{PatPet2003}.
Calculated best-fit parameters $\sigma$, $\alpha$, and $\beta$ of the in-medium pion scattering amplitude and respective $\chi^2$ values are given in the Table 1.

It is seen that our results are in a reasonable agreement with experimental data.
Some dissimilarity is observed only at large angles (discussed below).

Figure \ref{inmedium} shows the averaged values $X = (X_{\pi^+} + X_{\pi^-})/2$ where $X=\sigma,\alpha,\beta$ for the ``free''  $\pi^{\pm}N$-scattering parameters from \cite{Locher} in comparison with the obtained, also averaged, best-fit ``in-medium'' parameters  in dependence on $T^{lab}$.

Note, the bell-like forms of $\sigma^{free}$ and $\sigma^{eff}(T^{lab})$ have maximum at the same $T^{lab}$.
The dark gray (blue) domain $\sigma^{eff}$ is located below the light gray (yellow) $\sigma^{free}$ region.
This means that the ``in-medium'' $\pi^{\pm}N$-interaction becomes weaker as compared with that for ``free'' $\pi^{\pm}N$-scattering.

``In-medium'' $\alpha^{eff}(T^{lab})$ behavior indicates that refraction increases  at energy $T^{lab}>T_{res}^{lab}\simeq 170$ MeV.
It can be seen also that dark gray (blue) and light gray (yellow) regions become closer at  $T^{lab}>250$ MeV.

In our study we met two numerical problems which should be accounted for in future investigations.
First problem is already mentioned disagreement between calculated and experimental cross sections at large angles, and the visible dissimilarities increased with decreasing the energy.
This effect can be explained by the fact that the standardly applied Gaussian form of $\pi N$ form factor $f_{\pi}$ (see Eg.(\ref{MOP})
is not realistic in the region of large angles. Indeed, as experimentally established in \cite{Roper} the pion-nucleon cross section does not follow down but increases at angles over 80-100 degrees. Our calculation shows that agreement with experimental data is improved as we remove, in our fitting procedure, a few experimental points at large angles. It is demonstrated on Figure \ref{fig80}(a) for the case of $\pi^++^{28}$Si scattering at 180 MeV.

The other remark that should be pointed here is an ambiguity problem arising because the $\chi^2$ function (\ref{Nev}) has more than one minimum in the region of physically realistic parameters. In some cases two minima provide almost the same agreement with experimental data and additional information (such as total reaction cross sections) is needed to make a choice. This is demonstrated on the Figure \ref{fig80}(b) for the case $\pi^-+^{28}$Si at 130 MeV.

\section{Summary}
\begin{itemize}
\itemsep -1mm
\item
We show that the HEA-based three-parametric microscopic OP provides a reasonable agreement with experimental data of pion-nucleus elastic scattering at intermediate energies between 130 and 290 MeV.
\item
 Comparison of $\sigma^{free}$ and $\sigma^{eff}$ shows that,  at (3~3)-resonance energies, the $\pi N$-interaction in nuclear matter is weaker than in the case of free $\pi N$ collisions.
\item
Behavior of parameter $\alpha$  indicates that the refraction increases at energies more than $T^{lab}_{res}\simeq 170$ MeV.
\item
The decrease of the inmedium slope parameter $\beta^{eff}$ in comparison to the free one $\beta^{free}$ means that effective {\it rms} radius of the $\pi N$-system in nuclear medium becomes less than in the pion collisions with free nucleons
\item
Total cross section data are desirable to be involved to resolve the ambiguity problem.
\item
We should note that the usage of isotopically averaged parameters of $\pi^\pm N$-scattering amplitudes in the microscopic OP (\ref{MOP}) is available for nuclei with the same numbers of protons and neutrons $Z \simeq A-Z$ \cite{pion_izvran}. Hence the case of $\pi$-scattering  on $^{208}$Pb with significant difference between numbers of protons and neutrons  requires a special consideration.
\end{itemize}

\section*{Acknowledgements}
The work was partly supported by the Program ``JINR -- Bulgaria''. Authors E.V.Z. and K.V.L. thank the RFBR for the partial financial support under grant No. 12-01-00396a.

\end{document}